\documentclass[11pt]{article}
\usepackage{moriond,epsfig}

\def\be{\begin{equation}}
\def\ee{\end{equation}}
\def\bea{\begin{eqnarray}}
\def\eea{\end{eqnarray}}

\begin{document}
\vspace*{4cm}
\title{UNIVERSALITY IN MULTIFIELD INFLATION FROM STRING THEORY}

\author{ S. RENAUX-PETEL }

\address{Department of Applied Mathematics and Theoretical Physics,\\
University of Cambridge, Wilberforce Road, Cambridge, U.K.}

\maketitle\abstracts{
We develop a numerical statistical method to study linear cosmological fluctuations in inflationary scenarios with multiple fields, and apply it to an ensemble of six-field inflection point models in string theory. The latter are concrete microphysical realizations of quasi-single-field inflation, in which scalar masses are of order the Hubble parameter and an adiabatic limit is reached before the end of inflation. We find that slow-roll violations, bending trajectories and ``many-field'' effects are commonplace in realizations yielding more than 60 e-folds of inflation, although models in which these effects are substantial are in tension with observational constraints on the tilt of the scalar power spectrum.}

\section{Motivations and methodology}

Inflation provides both a beautiful mechanism to solve the conceptual problems of the Hot Big-Bang model as well as an elegant explanation for the observed spectrum of cosmic microwave background anisotropies. The simplest models of inflation, involving a single field slowly rolling down its potential, generate primordial fluctuations in agreement with observational requirements: adiabatic, Gaussian and nearly scale-invariant. However, more complicated models involving multiple light fields and/or violations of slow roll are more natural from a theoretical point of view. In particular, generic scalar fields during inflation acquire masses of order the Hubble parameter $H$ through gravitational-strength interactions, and therefore do not decouple from the inflationary dynamics. This general picture is well-attested in flux compactifications of string theory, in which lots of moduli are found with masses clustered around $H$. In cases where the moduli potential is computable, one generally finds a complicated, high-dimensional potential energy landscape with structure dictated by the spectrum of Planck-suppressed operators in the theory \cite{Baumann08}. The nature of inflation in such a potential is the important problem that we address in McAllister \textit{et al.} \cite{McAllister:2012am}.

For definiteness, we study the primordial fluctuations generated in a class of six-field inflationary models in string theory, corresponding to a D3-brane moving in a conifold region of a stabilized compactification. We draw potentials at random from a well-specified ensemble and study realizations that inflate by chance (they add up to more than 18400). We build on prior work \cite{Agarwal:2011wm} characterizing the homogeneous background evolution in this system. There, it was shown that inflationary trajectories lasting more than 60 e-folds take a characteristic form: the D3-brane initially moves rapidly in the angular directions of the conifold, spirals down to an inflection point in the potential, and then settles into an inflating phase. It was also established that the inflationary phenomenology has negligible dependence on the detailed form of the statistical distribution for the coefficients characterizing the potentials: a sort of universality emerges in this complicated ensemble. 

\section{Mass spectrum and the adiabatic limit}

A important aspect of our system is its mass spectrum, \textit{i.e.} the distribution of the 6 eigenvalues of the mass matrix of cosmological perturbations. Details of it, in particular the splitting between the various mass eigenvalues, can be understood using a random mass matrix model of supergravity theories \cite{McAllister:2012am,MMW}. We refer the reader to these papers for more details, and only emphasize here its main characteristic and consequences: except the lightest direction, which can happen to be very light ($ |m/H  | \ll 1$)) because of accidental cancellations, all other directions have masses of order $H$ (typical values are between $H$ and $5H$). This implies both that each direction is light enough to fluctuate and have interesting observational consequences, and that an ``adiabatic limit'' is naturally reached before the end of inflation.
To make this statement more precise, let us recall that cosmological fluctuations during $N$-field inflation are commonly divided into one instantaneous adiabatic perturbation and $N-1$ entropic perturbations, defined as fluctuations pointing respectively along and off the background trajectory \cite{Gordon:2000hv,GrootNibbelink:2001qt}. The key-feature of inflation with multiple fields lies in the fact that entropic perturbations, absent in single-field models, can feed the adiabatic perturbation, even on super-Hubble scales, when the trajectory bends. As we will explain, we observe that such effects have to be taken into account in our system and can have a profound impact on observable predictions. However, it turns out that the entropic perturbations, with typical masses greater than $H$, gets gradually diluted after Hubble-crossing until becoming completely ``exhausted'' by the end of inflation, with an amplitude suppressed by several order of magnitudes (typically more than 10) compared to the adiabatic perturbation. This dynamical process implies that that the curvature perturbation, proportional to the adiabatic one, has becomes constant by the end of inflation. It therefore allows us to make definite predictions for observables without having to make a detailed description of the reheating scenario, as well as it guarantees the absence of remaining primordial isocurvature perturbations in the radiation era \footnote{Barring an exponential growth of entropic fluctuations during violent (p)reheating processes.}. Put in another way, our ensemble provides concrete microphysical realizations of quasi-single-field inflation \cite{Chen:2009zp}, with the notable difference that self-couplings in our potentials are not large enough to generate observable non-Gaussianities \cite{McAllister:2012am}.

\section{Results on linear cosmological perturbations}

For each of our inflationary realizations, we numerically solve the exact dynamics of linear perturbations for the pivot scale crossing the Hubble radius 60 e-folds before the end of inflation, imposing Bunch-Davies initial conditions. We deduce the corresponding amplitude ${\cal P}$ and spectral index of the curvature perturbation when it has become constant at the end of inflation. Eventually, to disentangle the different physical effects affecting the results, we cross-correlate the exact results with some relevant background quantities and the predictions of three approximate models of the cosmological perturbations: the {\sf naive} (or slow-roll) model, which neglects multifield effects and assumes all fields are very light, gives the simple estimate ${\cal P_{{\rm \sf naive}}}=\frac{H_\star^2}{8 \pi^2 \epsilon_{\star}}$ where the star denotes evaluation at Hubble crossing. In the {\sf one-field} model, all entropic perturbations are set to zero but the full dynamics of the instantaneous adiabatic fluctuation is taken into account, including possible slow-roll violations. Finally, the {\sf two-field} model additionally takes into account one particularly picked entropic mode (see below).

We observe that the amplitude of the curvature power spectrum generically differs significantly from its naive estimate. This should not come as a surprise as, even neglecting entropic perturbations, the adiabatic mass at Hubble crossing is typically of order $H$, implying a smaller amplitude of large-scale fluctuations than what the naive estimate suggests. The {\sf one-field} power-spectrum ${\cal P}_1$, however, accurately describes, to the percent-level, 70 \% of our realizations. These cases, that we call effectively one-field, generally have a complicated non slow-roll dynamics, but nonetheless have negligible multiple field effects. In the following, we concentrate on the other substantial fraction of our realizations, about 30 \%, which are effectively multifield, \textit{i.e.} in which entropic perturbations can not be neglected.

We find that the impact of entropic perturbations is modest in the majority of these effectively multifield models, the ratio ${\cal P}/{\cal P}_{1}$ being less than 2 in about 65 \% of them, but that there exists a long tail toward large multifield effects, 15 \% (resp. 5 \%) of them having ${\cal P}/{\cal P}_{1}>10$ (resp. $>100$): \textit{i.e.} models in which entropic perturbations affect the amplitude of the power spectrum by one or several orders of magnitude are not rare. One should also note that, due to our specific typical background evolution, the amplitude of the multifield effects is correlated with the duration of inflation: models with a large number of e-folds of expansion ($>100$ for definiteness) generally already have reached their inflection point attractor 60 e-folds before the end of inflation, implying a modest degree of bending when our pivot scale is super-Hubble, and hence negligible conversion of entropic fluctuations into the curvature perturbation. Models with a shorter number of efolds, on the contrary, usually exhibit some bending around and after Hubble-crossing as remnants of their initial conditions, and are more liable to multifield effects. An even more remarkable fact though, which we believe has not been observed before, is the existence of a critical threshold of total turning (between Hubble-crossing and the end of inflation) to obtain a given amplitude of multifield effects, as measured by ${\cal P}/{\cal P}_{1}$ (see McAllister  \textit{et al.} \cite{McAllister:2012am} for more details).

In a general $N$-field model of inflation, one of the $N-1$ entropic fluctuation is usually picked: the one that instantaneously couples to the adiabatic perturbation \cite{GrootNibbelink:2001qt}, which we call the ``first'' entropic fluctuation. Most explicit studies of inflation with multiple fields have considered by simplicity 2-field models only, in which case the entropic subspace is one-dimensional and this distinction is unnecessary. In our system however, it is legitimate to assess the importance of \emph{many}-field effects, \textit{i.e.} to discriminate which models, amongst the effectively multifield ones, can be accurately described by the adiabatic and first entropic fluctuations only, which we call effectively two-field, and the ones in which the ``higher-order'' entropic modes affect the curvature perturbation, which we call effectively \emph{many}-field. For that purpose, we compare the {\sf exact} and the {\sf one-field} power spectrum to the one in which all these higher-order entropic modes are set to zero. Our results are as follows: we find that non-negligible many-field effects are commonplace, with only 38\% of effectively multifield models being effectively 2-field. There is also a clear pattern regarding the relative importance of 2-field versus many-field effects. We find that when multifield effects are modest in size, they are most often 2-field only. When they are large however, we find either large many-field effects, or large 2-field effects (in smaller proportion), but few models display both significant 2-field and many-field effects.

The above analysis was carried out without conditioning on obtaining a scalar spectral index in the observational window $0.93< n_s< 0.99$ (95\% CL) \cite{Larson:2010gs}. Once we impose this condition, we find that the overwhelming majority of viable realizations are effectively single-field and with small departures from slow-roll. In view of this, one might be inclined to brush off our findings as uninteresting. At least two reasons make us resist this inclination. First: although they are rare, there do exist models that would be naively ruled out from the naive estimate $n_s-1=-2 \epsilon_\star - \eta_\star$ and that turn out to be consistent with the observational constraints because of large multifield effects. Second and more important: we find that non-negligible multifield effects most often (more than 80\%) tend to reduce the spectral index significantly, by several factors of 0.1. This interesting and unexpected feature simply turns out to be not large enough in general to counterbalance the fact that the one-field spectral index is way too blue in the models with small duration of inflation which display significant bending in their last 60 e-folds. We find this trend nonetheless remarkable, and we think it leaves the interesting possibility that other classes of models, different from inflection point inflation, may generically display large multifield effects while meeting observational requirements.

\section{Conclusions and perspectives}

We have developed a statistical approach to the study of inflationary scenarios with multiple fields: it consists in comparing, in a large ensemble of realizations of a class of inflationary models, the {\sf exact} dynamics of linear fluctuations to three approximate descriptions: {\sf slow-roll}, {\sf single-field}, and {\sf two-field}, enabling us to characterize to which extent these simpler effective models are able to describe the physics of our class of models. We have applied this method to inflection point inflationary models with multiple fields of masses of order the Hubble parameter, as expected in generic low-energy effective field theories. We have observed that this mass spectrum implies that cosmological perturbations dynamically reach an ``adiabatic limit'' by the end of inflation, thereby alleviating the need for a precise description of reheating. We found that slow-roll violations and strongly bending trajectories are common in realizations yielding more than 60 e-folds of inflation, although the majority of models with an observably acceptable spectral index turns out to be effectively slow roll single-field. We think however that this pessimistic conclusion regarding the likelihood of models displaying large multifield effects and consistent with observations should be tempered, as it might be particular to inflection point inflation, and as we found that multifield effects most often tend to significantly redden the spectrum. We also demonstrated the existence of a critical threshold of turning to generate a given amplitude of multifield effects. Eventually, we pointed out the generic importance on cosmological perturbations of \textit{many} (beyond 2)-field effects, implying that the prevailing trend to study 2-field models of inflation may well be misleading to unveil the true nature of inflation with multiple fields. More generally, while a great deal of attention has been recently devoted to the study of the primordial non-Gaussianities generated during inflation, we think our study shows that many aspects of the \textit{linear} dynamics of cosmological perturbations during inflation with multiple fields remain to be explored, and we hope our work and our methodology will pave the way for further developments in this direction.

\section*{Acknowledgments}
I am supported by the STFC grant ST/F002998/1 and the Centre for Theoretical Cosmology. I would also like to thank the scientific committee of the Moriond conference for their financial support.


\section*{References}
\begin{thebibliography}{99}


\bibitem{Baumann08}
  D.~Baumann, A.~Dymarsky, S.~Kachru, I.~R.~Klebanov and L.~McAllister,
  JHEP {\bf 0903}, 093 (2009)
  
\bibitem{McAllister:2012am}
  L.~McAllister, S.~Renaux-Petel and G.~Xu,
  arXiv:1207.0317 [astro-ph.CO].

\bibitem{Agarwal:2011wm}N. Agarwal, R. Bean, L. McAllister, G. Xu, JCAP {\bf 1109} (2011) 002.

\bibitem{MMW}
  D.~Marsh, L.~McAllister and T.~Wrase,
  JHEP {\bf 1203}, 102 (2012)
  
\bibitem{Gordon:2000hv}
  C.~Gordon, D.~Wands, B.~A.~Bassett and R.~Maartens,
  Phys.\ Rev.\ D {\bf 63} (2001) 023506
  
\bibitem{GrootNibbelink:2001qt}
  S.~Groot Nibbelink and B.~J.~W.~van Tent,
  CQG {\bf 19} (2002) 613
  
\bibitem{Chen:2009zp}
  X.~Chen and Y.~Wang,
  JCAP {\bf 1004} (2010) 027
  
\bibitem{Larson:2010gs}
  D.~Larson {\it et al.},
  Astrophys.\ J.\ Suppl.\  {\bf 192} (2011) 16

\end{thebibliography}
\end{document}